\documentclass[twocolumn,superscriptaddress,aps,prb,showpacs,floatfix]{revtex4}

\usepackage{amsmath}
\usepackage{amssymb}
\usepackage{amsfonts}
\usepackage{pifont}
\usepackage{colordvi}
\usepackage{bm}
\usepackage{bbm}
\usepackage{epsf}
\usepackage{epsfig}
\usepackage{times}
\usepackage{nicefrac}
\usepackage{colordvi}
\usepackage{color}
\usepackage{graphicx}
\usepackage{dcolumn}

\def\sprm#1#2{  \left\langle #1 \left\vert \right. #2 \right\rangle   }
\def\mem#1#2#3{  \left\langle #1 \left\vert  #2 \right\vert #3 \right\rangle   }

\begin{document}

\preprint{Version 3.1}

\title{Quantum Dot Potentials: Symanzik Scaling, 
Resurgent Expansions and Quantum Dynamics}

\author{Andrey Surzhykov}
\affiliation{Max--Planck--Institut f\"ur Kernphysik,
Saupfercheckweg 1, 69117 Heidelberg, Germany}

\author{Michael Lubasch}
\affiliation{Max--Planck--Institut f\"ur Kernphysik,
Saupfercheckweg 1, 69117 Heidelberg, Germany}

\author{Jean Zinn--Justin}
\affiliation{DAPNIA, Commissariat \`{a} l'\'{E}nergie Atomique,
Centre de Saclay, 91191 Gif--Sur--Yvette, France}

\author{Ulrich D.~Jentschura\footnote{e-mail: ulj@mpi-hd.mpg.de}$^{,}$}
\affiliation{Max--Planck--Institut f\"ur Kernphysik,
Saupfercheckweg 1, 69117 Heidelberg, Germany}

\date{\today}

\begin{abstract}
This article is concerned with a special class of the ``double--well--like'' 
potentials that occur naturally in the analysis of finite
quantum systems. Special attention is paid, in particular, to 
the so--called Fokker--Planck potential, which has a particular property:
the perturbation series for the ground--state energy vanishes to all orders 
in the coupling parameter, but the actual ground-state energy
is positive and dominated by instanton configurations of the form 
$\exp(-a/g)$, where $a$ is the instanton action. The instanton
effects are most naturally taken into account within 
the modified Bohr--Sommerfeld quantization conditions whose 
expansion leads to the generalized perturbative expansions 
(so-called resurgent expansions) for the energy eigenvalues of the 
Fokker--Planck potential. Until now, these
resurgent expansions have been mainly applied for small values
of coupling parameter $g$, while much less attention has been paid to
the strong-coupling regime. In this contribution, we compare the 
energy values, obtained by directly resumming 
generalized Bohr--Sommerfeld quantization conditions, to the strong-coupling 
expansion, for which we determine the first few expansion coefficients 
in powers of $g^{-2/3}$. Detailed calculations are performed
for a wide range of coupling parameters $g$ and indicate a considerable 
overlap between the regions of validity of the 
weak-coupling resurgent series and 
of the strong--coupling expansion. Apart from the analysis of the 
energy spectrum of the Fokker--Planck Hamiltonian, we also briefly 
discuss the computation of its eigenfunctions. These
eigenfunctions may be utilized for the numerical integration of the 
(single-particle) time-dependent Schr\"odinger equation and, hence, for 
studying the dynamical evolution of the
wavepackets in the double--well--like potentials.     
\end{abstract}


\maketitle


\section{Introduction}
\label{intro}

In this paper, we study the all-order summation of instanton contributions
to the energy eigenvalues of anharmonic quantum mechanical oscillators
which involve (almost) degenerate minima.
The Euclidean path integral of quantum mechanical systems of
this kind with one space-
and one time-dimension is dominated by instanton configurations whose
action remains finite in the limit of large Euclidean (imaginary) times.
In order to find the energy eigenvalues, instanton configurations
have to be taken into account.
Modified quantization conditions have been conjectured for various
classes of potentials (for a review see
Refs.~\onlinecite{ZJJe2004i,ZJJe2004ii}), 
and accurate numerical calculations have been
verified against analytic expansions in the regime of small 
coupling~\cite{JeZJ2001,JeZJ2004plb}.
Indeed, for small coupling, the energy eigenvalues are dominated
by one-, two-, and three-instanton effects which correspond to
trajectories of the classical particle with
a small number of oscillations between the (almost) degenerate minima
of the potential. Here, we are concerned with the all-order
summation of the instanton contributions, which is applicable
to intermediate and strong coupling. Also, the transition
from small to strong coupling, and strong-coupling expansions,
will be discussed. Finally, we consider applications in 
the quantum dynamical simulation of finite systems.

The purpose of this paper thus is threefold:
first, to find generalized perturbative expansions
(so-called ``resurgent expansions'') for excited states
of certain classes of notoriously problematic~\cite{HeSi1978} 
quantum mechanical potentials, second, to derive large-coupling
asymptotics for these potentials and to investigate
overlap regions between small- and large-coupling asymptotics,
and third, to outline applications of the considerations
for the quantum dynamical simulation of a single particle 
in a double-well-like potential.
The first of these purposes is connected with mathematical physics,
the second is tied to Symanzik scaling in the 
``poor man's'' variant and therefore, to a basic implementation 
of the renormalization group, and the third one is 
rather application-oriented.

Within the first and second aims of our investigation, 
we also investigate the fundamental question 
whether small-coupling perturbative expansions 
can be continued analytically to the regime of 
large coupling, if one includes instantons into the 
formalism. Instantons can be considered 
either on the level of a resurgent expansion,
augmented by an optimized (generalized) Borel-Pad\'{e} resummation,
or on the level of a generalized quantization condition,
which entails an all-order resummation of the 
instanton expansion. 

The third application is mainly tied to the semiconductor 
``double quantum dot'' structures~\cite{LuEtAl1989,LoDV1998,HoEtAl2002,HuEtAl2005}, 
which are formed from two quantum dots coupled by quantum 
mechanical tunneling. Nowadays, these 
structures are generally accepted to belong to one of the most hopeful 
candidates for the realization of quantum bits (qubits), because
a single electron state in a double-well potential obviously can 
be localized in either of the two wells and, in that sense, represents a 
two-quantum-state system needed for quantum computing. 
Indeed, the theoretical analysis of the
structural and the dynamical properties of such (single-electron) 
double quantum dots can be traced back to double-well-like 
potentials. In this context, quantum dynamical calculations for a 
tunneling of a single particle between the two localized 
states nowadays attract special interest \cite{GrDiJuHa1991, Op1999} but 
require a detailed knowledge on the eigenvalues and eigenfunctions 
of the (double-well-potentials) potentials for a wide range
of coupling parameters. 

Another question may as well be asked: The effective 
instanton-related expansion parameter,
which reads $\Xi_1(g) = \sqrt{2/\pi}\, {\rm e}^{-1/6g}/g$
for the first two excited states of the Fokker--Planck
potential (as discussed below), is nonperturbatively
small for $g \to 0$, but numerically not very
small for some very moderate $g$. Specifically,
$\Xi_1(g)$ reaches its maximum $\Xi_1(1/6) = 1.76115\dots$ 
already at a rather small coupling parameter $g = 1/6$.
So, one may ask how the ``instanton expansion''
in powers of $\Xi_1(g)$ should be resummed, in addition to the
perturbative expansions about each instanton.
This latter step has never been accomplished, and we pursue its completion
via a direct resummation of generalized quantization conditions.

This paper is organized as follows.
In Sec.~\ref{frame}, basic definitions
related to the Fokker--Planck and the double-well
potential are recalled. Calculations are described
in Sec.~\ref{resumm}. Specifically, we consider
the resummation of the resurgent expansion
in Sec.~\ref{resurg}, the resummation of the 
quantization condition in Sec.~\ref{resumq}, 
large-coupling asymptotics in Sec.~\ref{largec},
and quantum dynamic simulations in Sec.~\ref{quadyn}.
Conclusions are drawn in Sec.~\ref{conclu}.

%
%
\section{Basic framework and numerical procedure}
\label{frame}

%
%
\subsection{Basic formulation}
\label{basic}

In this manuscript, we discuss the determination of the eigenvalues of the 
one--dimension Fokker--Planck (FP) Hamiltonian  
\begin{equation}
\label{hamfp}
H_{\rm FP} = -\frac{1}{2} \left( \frac{d}{dq} \right)^2 \, + \,
\frac{1}{2} q^2 \left( 1 - \sqrt{g}q \right)^2 \, + \, \sqrt{g} q \, - \, 
\frac{1}{2} \,,
\end{equation}
where $g$ is a positive coupling constant. 
For $g \, = \, 0$, Eq.~(\ref{hamfp}) represents
the Hamiltonian of the quantum harmonic oscillator whose eigenvalues are
given by the well known formula $E^{(K)} \, = \, K$, where 
$K = 0, 1, 2,\dots$ is the ``principal'' quantum number. For 
nonvanishing coupling, in contrast, no 
closed-form analytic expressions have been derived so far,
and approximations have to be used (for a classification
of the Fokker--Planck Hamiltonian in terms of a SUSY
algebra, see App.~\ref{appa}). The usefulness
of the notation $K$ instead of $N$ will become clear
in the following. If one considers the operator 
$V(g) = \sqrt{g} \, q - \sqrt{g} \, q^3 + g \, q^4/2$ in (\ref{hamfp}) 
as a perturbation and formally applies 
Rayleigh--Schr\"odinger perturbative expansion
to the $K$th harmonic oscillator state, then one finds the following
result for the first terms,
\begin{equation}
\label{epert}
E_{\rm FP, pert}^{(K)}(g) = K - 3K^2 g - 
\left(17 \, K^3 + \frac{5}{2} \, K \right) g^2   + 
\mathcal{O}(g^3) \,.
\end{equation}
All coefficients up to order $g^{80}$ 
are available for download~\cite{JeHomeHD}.
This perturbation expansion~\cite{HeSi1978}
fails to reproduce the spectrum of the Hamiltonian 
(\ref{hamfp}) even qualitatively. For instance, while the 
true ground--state energy $E^{(K = 0)}_{\rm FP}$ is manifestly nonvanishing
and positive, the perturbation series (\ref{epert}),
for $K = 0$, vanishes identically to all orders in the coupling $g$
and is thus formally converging to a zero energy eigenvalue.
A generalization of perturbation theory is required, therefore, 
in order to correctly describe the physical properties of the 
Fokker--Planck Hamiltonian, including its energy spectrum.

A complete description 
of the eigenvalues of the Hamiltonian (\ref{hamfp}) 
has been proposed recently~\cite{JeZJ2004plb,ZJJe2004i,ZJJe2004ii} by 
using a generalized perturbation series involving instanton 
contributions. Since the concept of instantons in quantum mechanics has 
been presented in a number of places 
\cite{JeZJ2004plb,ZJJe2004i,ZJJe2004ii,ZJ1984jmp}, we may here restrict 
ourselves to a rather short account of the basic formulas. In the 
semi--classical framework, the eigenvalues of the Fokker--Planck 
Hamiltonian can be found by solving the generalized Bohr--Sommerfeld 
quantization condition \cite{JeZJ2004plb,ZJ1984jmp}
\begin{eqnarray}
\label{qcond}
\lefteqn{\frac{1}{\Gamma(-B_{\rm FP}(E, g)) \,\,
\Gamma(1-B_{\rm FP}(E, g))}} \nonumber\\[2ex]
& & + \, \left(-\frac{2}{g}\right)^{2B_{\rm FP}(E, g)} 
\frac{{\rm e}^{-A_{\rm FP}(E, g)}}{2 \pi} \, = \, 0 \,.
\end{eqnarray}
In this expression, the functions $B_{\rm FP}(E, g)$ and 
$A_{\rm FP}(E, g)$ determine the perturbative expansion and the 
perturbative expansion about the instantons, correspondingly. 

The evaluation of these functions in terms of series in variables $E$ and $g$ 
has been described in detail elsewhere \cite{ZJJe2004i,ZJJe2004ii,ZJ1984jmp} 
for rather general classes of 
potentials. In the particular case of the Fokker--Planck potential, for 
example, the function $B_{\rm FP}(E, g)$ has the following expansion
[see Eq.~(14a) of Ref.~\onlinecite{JeZJ2004plb}]:
\begin{eqnarray}
\label{B_expansion}
B_{\rm FP}(E, g) \, =  E + 3E^2g + 
\left(35E^3 + \frac{5}{2}E\right)g^2 + \mathcal{O}(g^3) \, . 
\end{eqnarray}
The function
$B_{\rm FP}(E, g)$ alone
defines the perturbation expansion (\ref{epert}) which 
can be easily found by inverting the equation $B_{\rm FP}(E, g) \, = \, 
K$. The instanton contributions to the eigenvalues of the Fokker--Planck 
Hamiltonian are described by the function
[see Eq.~(14b) of Ref.~\onlinecite{JeZJ2004plb}]:
\begin{eqnarray}
\label{A_expansion}
\lefteqn{A_{\rm FP}(E, g) \, = \, \frac{1}{3g} \, + \, 
\left( 17E^2 +\frac{5}{6}\right) \, g } \nonumber\\[2ex]
& & + \, \left(227E^3 + \frac{55}{2}E\right)g^2 \, + \, 
\mathcal{O}(g^3) \, . 
\end{eqnarray}
Extensive numerical checks of the generalized quantization condition 
(\ref{qcond}) and the expansions (\ref{B_expansion})--(\ref{A_expansion}) have 
recently been performed for the ground state of the Fokker--Planck potential 
in the weak coupling regime \cite{JeZJ2004plb}. However, to the best of our 
knowledge, a numerical verification of these formulae 
(i) for excited states and (ii) for the large values of $g$
is still missing. The numerical checks will be presented in Sec.~\ref{resumq}.

While, of course, the present work is mainly devoted to the investigation of 
the energies and the corresponding 
wave functions of the Fokker--Planck potential, we will 
also briefly recall the properties of the well--known double-well potential
which is characterized by the Hamiltonian
\begin{equation}
\label{hamdw}
H_{\rm dw} = -\frac{1}{2} \left( \frac{d}{dq} \right)^2 \, + \,
\frac{1}{2} q^2 \left( 1 - \sqrt{g}q \right)^2 \, .
\end{equation}
Moreover, for the analysis of the energy spectra of the Hamiltonians 
(\ref{hamfp}) and (\ref{hamdw}) it is very convenient to introduce 
the interpolating potential:
\begin{equation}
\label{hami}
H_{\rm I} = -\frac{1}{2} \left( \frac{d}{dq} \right)^2 \, + \,
\frac{1}{2} q^2 \left( 1 - \sqrt{g}q \right)^2 \, + 
\eta \left( \sqrt{g} q - \frac{1}{2} \right)\, ,
\end{equation}
which corresponds to the double-well potential if $\eta = 0$,
whereas $\eta = 1$ gives the Fokker--Planck case.

%
%
\subsection{Numerical calculation of eigenenergies}
\label{numer}

In order to numerically calculate energy eigenvalues of the 
Fokker--Planck and of the double--well potential,
it is sufficient to consider matrix elements of these
potentials in the basis of harmonic oscillator eigenfunctions,
and to perform matrix diagonalization in a large basis
spanned by harmonic oscillator eigenfunctions
(typically, a number of $\gtrsim 5000$ basis functions is sufficient
for all calculations reported in the current article).
One then observes the apparent convergence of the 
eigenenergies as the size of the basis is increased.
This procedure is numerically stable
provided one uses quadrupole precision (128-bit, 32 decimal figure)
arithmetic. 

\begin{figure}[t]
\begin{center}
\includegraphics[height=0.9\linewidth,angle=270]{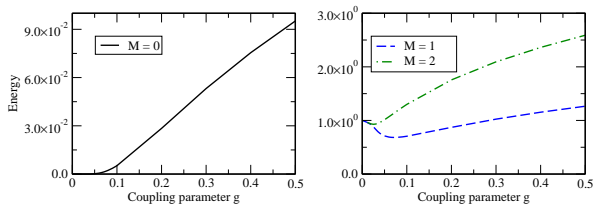}
\end{center}
\caption{\label{fig1} 
(Color online) Eigenvalues $E_{\rm FP}^{(M = 0)}$ (left panel) and 
$E_{\rm FP}^{(M = 1,2)}$ (right panel) of the Fokker--Planck Hamiltonian 
as a function of the coupling parameter $g$. Results have been computed by the
diagonalization of the Fokker--Planck Hamiltonian 
(\ref{hamfp}) in the basis of the harmonic 
oscillator wavefunctions.}
\end{figure}

Fokker--Planck energies of the lowest three eigenstates
[of the potential (\ref{hamfp})]
found by matrix diagonalization are displayed in Fig.~\ref{fig1} 
as a function of the coupling $g$ (we use the notation
$M = 0,1,2$ in order to denote these three energy levels). As seen 
from this Figure, the states $M = 1,2$ are degenerate
in the limit $g \to 0$. A similar energy level 
splitting is well known for the symmetric double--well potential 
\cite{ZJJe2004i,ZJJe2004ii,JeZJ2001} and may be explained in terms of 
nonperturbative instanton contributions. In contrast to the 
double--well case, the Fokker--Planck potential contains a linear 
symmetry--breaking term [cf. Eq.~(\ref{hamfp})], and this term
might be expected to lift any degeneracy. 
However, excited states can still develop a degeneracy
for $g \to 0$ in view of the (only perturbatively broken) parity 
$\varepsilon \, = \, \pm 1$ of the quantum eigenstates.
The ground state of the Fokker--Planck potential, however,
is located in one of the wells and does not develop
any degeneracy due to parity (see also Fig.~\ref{fig9} below).

\begin{figure}[ht]
\begin{center}
\includegraphics[width=0.7\linewidth,angle=270]{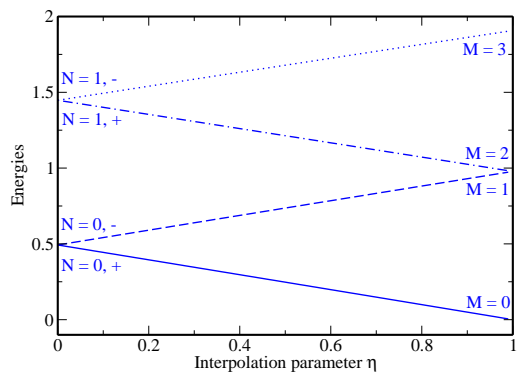}
\end{center}
\caption{\label{fig2} (Color online)
Adiabatic following of the lowest four eigenvalues 
of the interpolating potential (\ref{hami}) from $\eta = 0$
(double-well) to $\eta = 1$ (Fokker--Planck potential).
The value of the coupling parameter is held constant at
$g = 0.007$.}
\end{figure}

It is interesting to investigate the adiabatic following 
of eigenvalues for the interpolating potential (\ref{hami})
as a function of the parameter $\eta$ for fixed $g$.
This calculation (see Fig.~\ref{fig2}) reveals that the 
identification of the double-well 
energy eigenvalues~\cite{JeZJ2001,ZJJe2004i,ZJJe2004ii} with 
quantum numbers $(N,\varepsilon)$ for the double-well 
with the quantum number $M$
for the Fokker--Planck potential should proceed as follows:
\begin{subequations}
\label{ident}
\begin{align}
(N=0,+) \Leftrightarrow & \; M=0\,, \\
(N=0,-) \Leftrightarrow & \; M=1\,, \\
(N=1,+) \Leftrightarrow & \; M=2\,, \\
(N=1,-) \Leftrightarrow & \; M=3\,.
\end{align}
\end{subequations}
The general relation is $M = 2 N + (1 - \varepsilon)/2$.
However, the asymptotic behavior of the eigenenergies 
for $g \to 0$ is different in the two cases:
\begin{subequations}
\begin{align}
\label{NtoN}
E_{\rm dw}^{(N,\varepsilon)}(g) \to & \; N + \frac12\,, 
\qquad g \to 0 \,, \\
\label{MtoK}
E_{\rm FP}^{(M)}(g) \to & \; 
[\mkern - 2.5 mu  [ (M+1)/2 ] \mkern - 2.5 mu ] \,, 
\qquad g \to 0 \,,
\end{align}
\end{subequations}
where $[\mkern - 2.5 mu  [x] \mkern - 2.5 mu ]$ is the integral part of
$x$, i.e., the largest integer $m$ satisfying $m \le x$.
Equation (\ref{MtoK}) implies that the perturbative contribution
to the Fokker--Planck energy level with quantum number $M$ is 
given by Eq.~(\ref{epert}) with 
$K = [\mkern - 2.5 mu  [ (M+1)/2 ] \mkern - 2.5 mu ]$.

Apart from the degeneracy introduced by the instanton contributions, the 
eigenvalues $E^{(M = 1,2)}_{\rm FP}(g)$ also have a 
qualitatively different dependence on $g$ when compared to the 
ground--state energy $E^{(M = 0)}_{\rm FP}(g)$. As seen from Fig.~\ref{fig1}, while the 
energy $E^{(M = 0)}_{\rm FP}(g)$ increases monotonically as a function of the 
coupling constant $g$, the energies $E^{(M = 1,2)}_{\rm FP}(g)$ 
have minima at
$g_0 \, \approx \, 0.07$ and $g_0 \, \approx \, 0.025$, respectively.

%
%

%
%
\section{Resummations, Energy Eigenvalues and Quantum Dynamics}
\label{resumm}

%
%
\subsection{Resummation of the resurgent expansion}
\label{resurg}

The generalized Bohr--Sommerfeld quantization condition (\ref{qcond}) together 
with the expansions (\ref{B_expansion})---(\ref{A_expansion}) of the 
$A_{\rm FP}$ and $B_{\rm FP}$ functions uniquely determines the eigenvalues 
of the Fokker--Planck Hamiltonian. For the ground state,
the energy eigenvalue can be found by systematic expansion
of (\ref{qcond}) in powers of the two small parameters $\exp(-1/3g)$ and $g$,
whereas for excited states, the parameters are $\exp(-1/6g)$ and $g$.
Thus, for the Fokker--Planck potential, the particular form of the
expansion differs for the ground vs.~excited states.

Explicitly, the ground-state energy ($M=0$) is given by the resurgent
expansion \cite{JeZJ2004plb}: 
\begin{equation}
\label{eground}
E^{(0)}_{\rm FP}(g) =
\sum^{\infty}_{n=1} \left( {{\rm e}^{-1/3g} \over 2 \pi} \right)^{n} \,
\sum^{n-1}_{k=0} \left\{ \ln\left(-\frac{2}{g}\right) \right\}^k \,
\sum^{\infty}_{l=0} f^{(0)}_{nkl} \, g^{l} \, ,
\end{equation}
where the index $n$ denotes the order of the ``instanton contribution'':
$n = 1$ is a one-instanton, $n=2$ is a two-instanton, etc. 
(as noted below, the one-instanton configuration involves a 
back-tunneling of the particle to the lower well for the 
ground state and thus has twice the action of the characteristic 
one-instanton effect for excited states). Another subtle point which 
should be recalled here is that the leading one-instanton term 
involves a summation over all possible $n$--instanton configurations but
neglects instanton interactions \cite{ZJJe2004i}. As seen from 
Eq.~(\ref{eground}), the evaluation
of the ground state energy within such a (first--order) approximation, 
also referred as a ``dilute instanton gas'' approximation, requires the 
knowledge of the $f^{(0)}_{10l}$ coefficients. Since these coefficients are 
available for download~\cite{JeHomeHD}, we only recall the six leading 
ones~\cite{JeZJ2004plb}:
\begin{eqnarray}
f^{(0)}_{100} &=& 1 \,, \quad 
f^{(0)}_{101} = -\frac{5}{6} \,, \quad 
f^{(0)}_{102} = -\frac{155}{72} \,, \nonumber \\[1ex]
f^{(0)}_{103} &=& -\frac{17315}{1296} \,, \quad
f^{(0)}_{104} = -\frac{3924815}{31104} \,, \\[1ex]
f^{(0)}_{105} &=& -\frac{294332125}{186624} \,, \quad 
f^{(0)}_{106} = -\frac{163968231175}{6718464} \,. \nonumber
\end{eqnarray}
By inserting these coefficients in Eq.~(\ref{eground}), we are able
to perform now the numerical check of the validity of the one-instanton 
expansion for the ground state at small coupling. In Table~\ref{table1},
for example, the energy $E^{(0)}_{\rm FP}(g)$ is displayed for coupling 
parameters in the range 0.005 $\le g \le$ 0.03 and is compared to the 
``true'' eigenvalues as obtained from the diagonalization of the 
Fokker--Planck potential in the basis of harmonic oscillator wavefunctions.
As seen from Table~\ref{table1}, the ground state energy is 
dominated by the one-instanton effect for relatively small 
values of the coupling parameter $g \le$ 0.01. 
For stronger coupling, however,
large discrepancies between the ``true'' energies and the results of
resummation of Eq.~(\ref{eground}) at $n$ = 1 are found indicating 
the importance of the higher--instanton terms which take into account
the instanton interactions. The evaluation of the higher--order corrections 
($n \ge$ 2) to the ground--state energy $E^{(0)}_{\rm FP}(g)$ is, however, 
a very difficult task since it requires a double resummation of the resurgent 
expansion, in powers of both $g$ and $\exp(-1/3g)$. In the 
present work such a double summation based on sequentially
adding higher-order instanton terms will not be performed.
Still, for the sake of completeness, we here indicate the leading 
two--instanton \cite{JeZJ2004plb} and three--instanton coefficients
for the Fokker--Planck ground state:  
\begin{eqnarray}
f^{(0)}_{210} &=& 2 \,, \quad
f^{(0)}_{200} = 2 \gamma \,, \nonumber \\[1ex]
f^{(0)}_{211} &=& -\frac{10}{3} \,, \quad
f^{(0)}_{201} = -\frac{10}{3} \gamma - 3 \,, \\[1ex]
f^{(0)}_{310} &=& 8 \, \gamma \,, \quad
f^{(0)}_{300} = 6 \gamma^2 + \frac{\pi^2}{6}\,, \nonumber \\[1ex]
f^{(0)}_{311} &=& -\frac{80}{3} \gamma - 6 \,, \quad
f^{(0)}_{301} = -15 \gamma^2 - 12 \gamma -17 + \frac{5}{12} \pi^2 \,, \nonumber
\end{eqnarray}
where $\gamma = 0.577216\dots$ is Euler's constant. 

%
%
\begin{table}[t]
\caption{\label{table1} Ground--state energy of the Fokker--Plank 
Hamiltonian. Results have been computed by the diagonalization of the
Hamiltonian in the basis of the harmonic oscillator wavefunctions ("true" 
energy) and by resummation of Eq.~(\ref{eground}) within the one-instanton
approximation ($n$ = 1). The numerical uncertainty of the entries 
in the right column is estimated on the basis of the apparent 
convergence of results under an appropriate
increase of the number of $f^{(0)}_{10l}$ parameters, which corresponds
to the number of terms in the perturbative expansion about the leading 
instanton. Numerical discrepancies between the left and right column
are due to higher-order instanton contributions, as described in the text.
We underline those decimal figures in the one-instanton results 
which are equal to the corresponding ones in the complete 
numerical solution.}
\vspace*{0.3cm}
\begin{scriptsize}
\begin{center}
\begin{tabular}{l@{\hspace*{0.5cm}}l@{\hspace*{0.5cm}}l} 
\hline
\hline
\rule[-3mm]{0mm}{8mm}
$g$ & $E^{(M = 0)}_{\rm FP}$ (diagonalization) 
    & $n = 1$ term of Eq.~(\ref{eground}) \\
\hline
\rule[-3mm]{0mm}{8mm}
0.005 &             1.766 107 332 563      $\times$ 10$^{-30}$  &
         \underline{1.766 107 332 563}     $\times$ 10$^{-30}$ \\
\rule[-3mm]{0mm}{8mm}
0.010 &             5.267 473 259 637      $\times$ 10$^{-16}$ & 
         \underline{5.267 473 259 637}     $\times$ 10$^{-16}$ \\
\rule[-3mm]{0mm}{8mm}
0.015 &             3.508 587 565 372      $\times$ 10$^{-11}$  &
         \underline{3.508 587 56}4 030     $\times$ 10$^{-11}$ \\
\rule[-3mm]{0mm}{8mm}
0.020 &             9.033 155 571 641      $\times$ 10$^{-09}$  &
         \underline{9.033 15}4 730 920     $\times$ 10$^{-09}$ \\
\rule[-3mm]{0mm}{8mm}
0.025 &             2.519 767 018 258      $\times$ 10$^{-07}$  &
         \underline{2.519 76}0 770 755(1)  $\times$ 10$^{-07}$ \\
\rule[-3mm]{0mm}{8mm}
0.030 &             2.313 302 179 961      $\times$ 10$^{-06}$  &
         \underline{2.313} 251 574 075(2)  $\times$ 10$^{-06}$ \\
\hline
\hline
\end{tabular}
\end{center}
\end{scriptsize}
\end{table}

As seen from Eq.~(\ref{eground}), no splitting into levels with 
positive and negative parity arises for the ground state of the 
Fokker--Planck potential due to the linear symmetry--breaking term 
in Eq.~(\ref{hamfp}). This term modifies the potential in such a way that the
leading, one-instanton ($n$ = 1) shift of the ground state energy
results from a back--tunneling (instanton--antiinstanton configuration)    
of the particle to the lower well \cite{JeZJ2004plb}. For excited states, 
in contrast, the one-instanton configuration is a trajectory which starts in
one well and ends in the other, restoring the broken symmetry.
Therefore, any excited state ($M >$ 0) of the Fokker--Planck Hamiltonian can 
be characterized by its principal quantum number 
\begin{equation}
  \label{FP_M_to_K}
  K = [\mkern - 2.5 mu  [ (M+1)/2 ] \mkern - 2.5 mu ]
\end{equation} 
and the parity 
\begin{equation}
   \label{FP_M_to_epsilon}
   \varepsilon = 2 \left( 2 K - M - \frac12 \right) \, .
\end{equation}
In fact, this classification is very
similar to the double-well potential (\ref{hamdw}) except,
of course, the particular case of the ground state. It follows
naturally that the resurgent expansion for
the excited states of the Fokker--Planck potential is very close to the
analogous expansion for the double--well potential and reads 
\cite{JeZJ2004plb}:
\begin{align}
\label{efpgen}
& E^{(M > 0)}_{\rm FP}(g) = 
E^{(K, \varepsilon)}_{\rm FP}(g) =
\sum_{l = 0}^\infty E_{K, l} \, g^l 
\nonumber\\[2ex]
& + \sum^{\infty}_{n=1} \left[-\varepsilon \, \Xi_K(g) \right]^n \,
\sum^{n-1}_{k=0} \left\{ \ln\left(-\frac{2}{g}\right) \right\}^k \,
\sum^{\infty}_{l=0} f^{(K)}_{nkl} \, g^{l} \, ,
\end{align}
where $E_{K, l}$ are perturbative coefficients and
$\Xi_K(g)$ is given by
\begin{equation}
\Xi_K(g) = \frac{2^{K-1/2}}{g^K\, \sqrt{ \pi \, K! \, (K-1)!}}\,
\rm{e}^{ - 1/6 g} \, .
\end{equation}
The power of $\Xi$ can again 
be associated with the order of the instanton
($K = 1$: one-instanton, $K = 2$ means two-instanton, etc.).
One should note that two intricacies are associated to 
the precise meaning of the quantities that enter Eq.~(\ref{efpgen}):
\begin{itemize}
\item In analogy to the double-well potential, the
imaginary part which is generated by the resummation of the 
perturbation series about the leading instanton
(the ``discontinuity'' of the distributional Borel sum
in the terminology of Ref.~\onlinecite{CaGrMa1986})
is compensated by an explicit imaginary part that stems from the
two-instanton effect [from the factor $\ln(-2/g)$].
Related questions have been discussed at length in 
Refs.~\onlinecite{ZJJe2004i,ZJJe2004ii}.
\item In contrast to the ground--state energy (\ref{eground}), the leading
contribution to the 
energies $E^{(\varepsilon,K > 0)}_{\rm FP}(g)$ for small coupling
arises from the perturbation expansion (\ref{epert}) which is manifestly
nonvanishing to all orders in $g$. However, since this perturbation expansion
is independent of the parity $\varepsilon$, the energy splitting of 
the levels with the same principal quantum number $K$ is again dominated by the
one-instanton contribution ($n = 1$). 
\end{itemize}
Similar to the ground state 
(\ref{eground}), we may compute such a contribution and, hence, a
splitting of an arbitrary excited state $K$ by making use of 
the $f^{(K)}_{10l}$ coefficients, which for $K > 0$ read
($l = 0,1,2,3$): 
\begin{subequations}
\begin{align}
f^{(K)}_{100} =& 1 \,, \qquad
f^{(K)}_{101} =  -\frac{17}{2} K^2 - 6 K - \frac{5}{12} \,,
\\[1ex]
f^{(K)}_{102} =& \frac{289}{8} K^4
- \frac{23}{2} K^3 - \frac{1139}{24} K^2 - \frac{45}{4} K 
- \frac{695}{288} \,,
\\[1ex]
f^{(K)}_{103} =& -\frac{4913}{48} K^6 
+ \frac{629}{2} K^5 + \frac{1637}{32} K^4 \nonumber\\[2ex]
& - \frac{1885}{3} K^3 
- \frac{155825}{576} K^2 - \frac{3835}{24} K - \frac{68885}{10368} \,.
\end{align}
\end{subequations}
Results for $K \leq 28$ are available for download~\cite{JeHomeHD}.
In Table~\ref{table2}, for example, the splitting 
$E^{(\varepsilon = -1,K = 1)}_{\rm FP}(g) - 
E^{(\varepsilon = +1,K = 1)}_{\rm FP}(g)$ of the first 
two excited states $M = 1,2$ due to the one-instanton effect 
($n$ = 1) is displayed as 
a function of the coupling parameter $g$. Again, a comparison of the results 
obtained by the resummation of Eq.~(\ref{efpgen}) and by the diagonalization of 
the Fokker--Planck Hamiltonian indicates the importance of the 
higher--instanton
effects ($n >$ 1) and, hence, the necessity of a double resummation of the 
resurgent expansion, in powers of both $g$ and $\exp(-1/6g)$. Instead of
performing such a double summation explicitly, it is more
convenient to enter directly into the 
quantization condition (\ref{qcond}), with resummed 
quantities as defined by the 
$A_{\rm FP}(E, g)$ and $B_{\rm FP}(E, g)$ functions. We discuss this
alternative approach in the next Section.

%
%
\begin{table}[t]
\caption{\label{table2} Energy difference between the excited states
$E^{(M = 1, 2)}_{\rm FP}(g)$ of the Fokker--Planck 
Hamiltonian. Results have been computed by
diagonalizing the Hamiltonian in the basis of the harmonic oscillator 
wavefunctions (left column) and by resummation of Eq.~(\ref{efpgen}) within 
the one-instanton approximation (right column). The numerical uncertainty is
estimated on the basis of the apparent 
convergence of results under an appropriate
increase of the number of $f^{(K)}_{10l}$ parameters.
As in Table~\ref{table1},
we underline those decimal figures in the one-instanton results
which are equal to the corresponding ones in the complete 
numerical solution.}
\vspace*{0.3cm}
\begin{scriptsize}
\begin{center}
\begin{tabular}{l@{\hspace*{0.5cm}}l@{\hspace*{0.5cm}}l} 
\hline
\hline
\rule[-3mm]{0mm}{8mm}
$g$  & $E^{(M = 2)}_{\rm FP} - E^{(M = 1)}_{\rm FP}$ (diag.) 
     & $n = 1$ term of Eq.~(\ref{efpgen}) \\
\hline
\rule[-3mm]{0mm}{8mm}
0.005 &             9.848 553 978 903       $\times$ 10$^{-13}$  &
         \underline{9.848 553 978 903}      $\times$ 10$^{-13}$ \\
\rule[-3mm]{0mm}{8mm}
0.010 &             7.801 059 663 554       $\times$ 10$^{-06}$ & 
         \underline{7.801 059 65}9 99(1)    $\times$ 10$^{-06}$ \\
\rule[-3mm]{0mm}{8mm}
0.015 &             1.213 924 539 483       $\times$ 10$^{-03}$  &
         \underline{1.213 9}1 452(2)        $\times$ 10$^{-03}$ \\
\rule[-3mm]{0mm}{8mm}
0.020 &             1.289 613 568 640       $\times$ 10$^{-02}$  &
         \underline{1.28}8 765(1)           $\times$ 10$^{-02}$ \\
\rule[-3mm]{0mm}{8mm}
0.025 &             4.633 794 364 814       $\times$ 10$^{-02}$  &
         \underline{4.6}11 6(1)             $\times$ 10$^{-02}$ \\
\rule[-3mm]{0mm}{8mm}
0.030 &             9.699 341 140 782       $\times$ 10$^{-02}$  &
         \underline{9.6}10 2(6)             $\times$ 10$^{-02}$ \\
\hline
\hline
\end{tabular}
\end{center}
\end{scriptsize}
\end{table}
%
%
%

%
%
\subsection{Resummation of the quantization condition}
\label{resumq}

The resurgent expansions (\ref{eground}) and (\ref{efpgen}) for the energies of
the ground and excited states of the Fokker--Planck Hamiltonian
follow as a direct consequence of the quantization condition (\ref{qcond}).
As seen from our calculations summarized in Tables \ref{table1} and 
\ref{table2}, these expansions are very useful for small coupling,
but not of particular usefulness even for rather
moderate values of $g$, because of the necessity of their double resummation.
Here, we would like to investigate whether it is possible to 
resum the divergent series that gives rise to $A_{\rm FP}(E, g)$ and 
$B_{\rm FP}(E, g)$ directly and look for solutions of the 
quantization condition (\ref{qcond}) without any
intermediate recourse to the resurgent 
expansion. In fact, this approach currently
appears to be the only feasible way to evaluate the 
multi--instanton expansion (in powers of $n$),
because the quantization condition incorporates all instanton orders. 

In order to introduce such a ``direct summation'' approach, we recall
that the solution of the generalized Bohr--Sommerfeld quantization condition 
(\ref{qcond}) for a particular coupling parameter $g$ must provide the 
energy spectrum of the Fokker--Planck Hamiltonian. In other words, if 
one defines the left--hand side of the quantization condition 
(\ref{qcond}) as a function of two variables $E$ and $g$:
\begin{align}
\label{Q_function}
Q(E, g) =& \, 
\frac{1}{\Gamma(-B_{\rm FP}(E, g)) \, \Gamma(1-B_{\rm FP}(E, g))} 
\nonumber\\[1ex]
& + \left(-\frac{2}{g}\right)^{2B_{\rm FP}(E, g)} 
\frac{{\rm e}^{-A_{\rm FP}(E, g)}}{2 \pi} \,,
\end{align}
then the zeros of this function at fixed $g$ determine the energy spectrum of 
(\ref{hamfp}):
\begin{equation}
   \label{Q_zeros}
   Q(E^{(M)}_{\rm FP}(g), g) \, = \, 0, \qquad M = 0, 1, 2, .... \, .
\end{equation}
A numerical analysis of the function $Q(E, g)$ can
be used, therefore, in order to examine
the validity and applicability of the generalized
quantization condition given by
Eqs.~(\ref{qcond})---(\ref{A_expansion}) 
for the case of strong coupling.

\begin{figure}[t]
\begin{center}
\vspace*{-0.5cm}
\includegraphics[width=0.5\linewidth,angle=270]{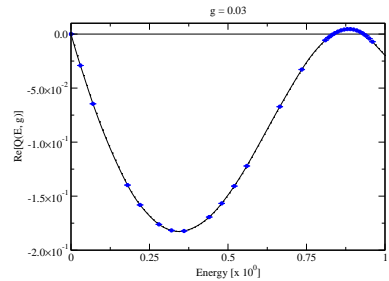}
\end{center}
\vspace*{-0.7cm}
\caption{\label{fig3} (Color online) 
Energy dependence of (the real part of) the 
function $Q(E, g)$. Calculations have been performed for 
fixed $g = 0.03$.}
\end{figure}

As seen from Eq.~(\ref{Q_function}), any analysis of the function 
$Q(E, g)$ can be traced back to the evaluation of the functions 
$A_{\rm FP}(E, g)$ and $B_{\rm FP}(E, g)$ which constitute series in two
variables, namely
$E$ and $g$ [cf. Eqs.~(\ref{B_expansion})---(\ref{A_expansion})].
In order to compute these series, it is convenient to re--write the 
functions $A_{\rm FP}(E, g) \equiv A_{\rm FP}(E, g x) |_{x=1}$ and 
$B_{\rm FP}(E, g) \equiv B_{\rm FP}(E, g x) |_{x=1}$ 
as (formal) power series in terms of a variable $x$, taken at $x$ = 1
[cf. \S 8.5 of Ref.~\onlinecite{Ha1949}]:
\begin{eqnarray}
\label{a_x_series}
B_{\rm FP}(E, g) \, = \, \sum\limits_{n = 0}^{N_{max}} 
\left. b^{(n)}_{\rm FP}(E, g) \, x^n \right|_{x=1}\, ,		
\end{eqnarray}
\begin{eqnarray}
\label{b_x_series}
A_{\rm FP}(E, g) - \frac{1}{3 g} \, = \, \sum\limits_{n = 0}^{N_{max}} 
\left. a^{(n)}_{\rm FP}(E, g) \, x^n \right|_{x=1}\, ,
\end{eqnarray}
where the coefficients 
$a^{(n)}_{\rm FP}(E, g)$ and $b^{(n)}_{\rm FP}(E, g)$ 
are uniquely determined by Eqs.~(\ref{B_expansion}) and (\ref{A_expansion}): 
$b^{(0)}_{\rm FP}(E, g) \, = \, E$, 
$b^{(1)}_{\rm FP}(E, g) \, = \, 3E^2g$,
etc. In the 
computations, the power series (\ref{a_x_series}) and (\ref{b_x_series}) 
allow one to use a unified computer algebra routine for the Borel--like 
summations, which simply takes as input the variables $a^{(n)}_{\rm FP}(E, g)$
and $b^{(n)}_{\rm FP}(E, g)$, as a function of $g$ and returns the
value of the resummed series at $x=1$. Indeed, this routine can be universally 
used for different values of $g$ and is therefore convenient for 
further numerical computations which are discussed below.   

Making use of Eqs.~(\ref{a_x_series}) and (\ref{b_x_series}), we 
may now perform a simultaneous summation of the perturbation series as 
well as of the perturbation series about each of the instantons and find the 
functions $B_{\rm FP}(E, g)$ and $A_{\rm FP}(E, g)$, correspondingly,
where we use the same notation for a function and its Borel sum.
There is a small subtlety because 
for positive $g$, the power series (\ref{a_x_series}) and 
(\ref{b_x_series}) are nonalternating and divergent and, hence, special 
resummation techniques are required to calculate the 
Borel sums. In our present 
calculations, for example, we apply a generalized Borel--Pad\'e 
method \cite{Je2000prd,JeSo2001}. The discussion of this method is
beyond the scope of the present work, and the reader is referred to 
Refs.~\onlinecite{FrGrSi1985,Je2000prd,JeSo2001} for a more detailed discussion.
Because of their nonalternating property,
the perturbation series defining the functions
$A_{\rm FP}(E, g)$ and $B_{\rm FP}(E, g)$ are Borel summable
only in the distributional sense~\cite{CaGrMa1986}.
The evaluation of the Borel--Laplace integral thus requires an
integration along a contour which is tilted with respect to the real
axis (for details see 
Refs.~\onlinecite{FrGrSi1985,Je2001pra,JeSo2001}
and the contours $C_{+1}$ and $C_{-1}$ in Ref.~\onlinecite{Je2000prd}).
The resummation of the divergent series 
(\ref{a_x_series}) and (\ref{b_x_series}) may be carried out along each of 
these contours, but it is important to characterize the perturbative and
instanton contributions in the same way, i.e. to deform the contours
for $B_{\rm FP}(E, g)$ and $A_{\rm FP}(E, g)$
either above or below the real axis, consistently. In the 
terminology of Ref.~\onlinecite{CaGrMa1986}, one should 
exclusively use either ``upper sums'' or ``lower sums,'' but
mixed prescriptions are forbidden. From a historical perspective, it is 
interesting to remark that the possibility of deforming the  
Borel integration contour had already been anticipated in
a remark near the end of Chap.~8 of the classic Ref.~\onlinecite{Ha1949}.

\begin{table}[t]
\caption{\label{table3} Ground--state energy of the Fokker--Plank 
Hamiltonian. Results have been computed by the diagonalization of the
Hamiltonian in the basis of the harmonic oscillator wavefunctions 
(left column) and by solving Eq.~(\ref{Q_zeros}), 
as indicated in the right column.}
\vspace*{0.3cm}
\begin{scriptsize}
\begin{center}
\begin{tabular}{l@{\hspace*{0.5cm}}l@{\hspace*{0.5cm}}l} 
\hline
\hline
\rule[-3mm]{0mm}{8mm}
$g$   & $E^{(0)}_{\rm FP}$ (diagonalization)& Zero of ${\rm Re}[Q(E, g)]$ \\
\hline
\rule[-3mm]{0mm}{8mm}
0.010 &             5.267 473 259 637     $\times$ 10$^{-16}$ & 
         \underline{5.267 473 259 637}    $\times$ 10$^{-16}$ \\
\rule[-3mm]{0mm}{8mm}
0.030 &             2.313 302 179 961     $\times$ 10$^{-06}$  &
         \underline{2.313 302 17}(2)      $\times$ 10$^{-06}$ \\
\rule[-3mm]{0mm}{8mm}
0.070 &             1.267 755 797 982     $\times$ 10$^{-03}$  &
         \underline{1.267 7}4(6)          $\times$ 10$^{-03}$ \\
\rule[-3mm]{0mm}{8mm}
0.100 &             5.199 138 696 222     $\times$ 10$^{-03}$ & 
         \underline{5.199} 3(2)           $\times$ 10$^{-03}$ \\
\rule[-3mm]{0mm}{8mm}
0.170 &             2.079 244 408 360     $\times$ 10$^{-02}$ & 
         \underline{2.07}8(1)             $\times$ 10$^{-02}$ \\
\rule[-3mm]{0mm}{8mm}
0.300 &             5.318 357 438 655     $\times$ 10$^{-02}$ & 
         \underline{5.3}23(9)             $\times$ 10$^{-02}$ \\
\hline
\hline
\end{tabular}
\end{center}
\end{scriptsize}
\end{table}

\begin{figure}[t]
\begin{center}
\includegraphics[height=0.9\linewidth,angle=270]{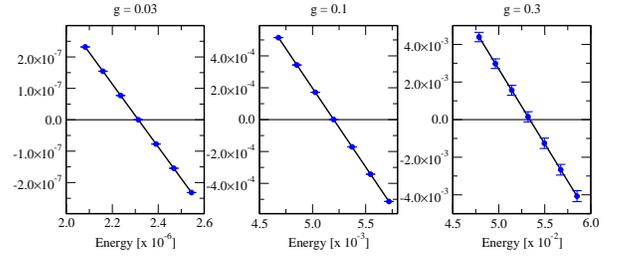}
\end{center}
\caption{\label{fig4} (Color online)
Energy dependence of (the real part of) the 
function $Q(E, g)$. As in Fig.~\ref{fig3},
the results of linear regression analysis of the 
function $Q(E, g)$ are depicted by the solid line. Calculations have been 
performed around the ``true'' energies of the ground state and for the 
different values of the coupling parameter: $g$ = 0.03 (left panel), 
$g = 0.1$ (middle panel) and $g = 0.3$ (right panel). The energy
eigenvalues obtained using the displayed graphs determine the 
corresponding entries in the right column of Table~\ref{table3}.}
\end{figure}

We are now in the position 
to analyze the properties of the function $Q(E, g)$ and,
hence, to extract the energy spectrum of the Fokker--Planck Hamiltonian.
As mentioned above, to perform such an analysis for any particular $g$ we have 
to (i) resum the (divergent) series for the functions $A_{\rm FP}(E, g)$ and 
$B_{\rm FP}(E, g)$ and (ii) insert the resulting generalized Borel sums 
into Eq.~(\ref{Q_function}). We may then interpret 
the $Q(E, g)$ as a function of 
$E$ (at fixed $g$) and (iii) numerically determine the zeros of this 
function which correspond to the energy values $E^{(M)}_{\rm FP}$ of the
Fokker--Plank Hamiltonian, according to Eq.~(\ref{Q_zeros}).   
In Fig.~\ref{fig3}, for instance, we display the energy dependence of the
real part of the function $Q(E, g)$ taken at $g = 0.03$. 
In the energy range $0 < E < 1$, this function has three zeros which
obviously correspond to the ground $E^{(0)}_{\rm FP}$ and to the excited 
$E^{(1, 2)}_{\rm FP}$ states. The ground-state energy 
$E^{(0)}_{\rm FP}$ determined in such a way is presented in Table~\ref{table3} 
and compared to reference values obtained by the diagonalization of the
Hamiltonian matrix in the basis of harmonic oscillator wavefunctions. Moreover, 
apart from the particular case of $g = 0.03$, we also display 
the energy $E^{(0)}_{\rm FP}$ for other coupling parameters spanning 
the range from $g = 0.01$ to $g = 0.3$ (see also Fig.~\ref{fig4}).
This rather wide range of coupling parameters $g$ considered here allows us to 
investigate the behavior of the generalized Bohr--Sommerfeld quantization
condition in the transition from weak to strong coupling. 
As seen from Table~\ref{table3}, the ground--state energy is well reproduced 
at $g = 0.01$ (up to 14 decimal digits).
Alternatively, a highly accurate value of the ground 
state energy (at $g = 0.01$) can be obtained from the  
the one-instanton contribution to the resurgent expansion (\ref{eground})
for $g < 0.01$, as indicated in Table~\ref{table1}. 
The accuracy of the one-instanton approximation is 
rapidly decreasing for higher $g$. For instance, at the moderate value of 
$g = 0.03$, the one-instanton term of the 
resurgent expansion (\ref{eground}) reproduces the ground-state
energy only to four decimal digits (see the last row of 
Table~\ref{table1}), while a 
total of eight digits can be obtained from solving 
Eq.~(\ref{Q_zeros}), as indicated in the second row of Table~\ref{table3}. 
For even stronger coupling, one observes a much larger 
numerical uncertainty in the determination of the zeros of the 
function $Q(E, g)$, because the convergence of the 
generalized and optimized Borel--Pad\'{e} methods employed in 
the resummation of the 
$A_{\rm FP}(E, g)$ and $B_{\rm FP}(E, g)$ functions 
is empirically observed to reach fundamental
limits for larger values of the coupling, which cannot be overcome
by the use of multiprecision arithmetic and might indicate a
fundamental limitation for the convergence of the transforms
and are not due to numerical cancellations. It might be interesting
to explore these limits also from a mathematical point of view.
Specifically, we have determined the numerical uncertainty
of the $A_{\rm FP}(E, g)$ and $B_{\rm FP}(E, g)$ functions 
on the basis of the apparent convergence of the Borel--Pad\'{e}
approximants, integrated in the complex plane and accelerated 
according to Ref.~\onlinecite{JeSo2001}, using an optimal truncation
of the order of the transforms. We found that as the order of
the Borel--Pad\'{e} transformation was increased, the apparent 
convergence of the transforms stopped at around order 40 for 
$g = 0.03$ and higher. Despite these difficulties, the generalized 
Bohr--Sommerfeld quantization formula (\ref{Q_zeros})
determines the ground--state energy of the 
Fokker--Planck potential with an accuracy of about 0.01 \% 
up to $g \leq 0.3$ (cf. Table~\ref{table3}).

\begin{figure}[t]
\begin{center}
\includegraphics[height=0.9\linewidth,angle=270]{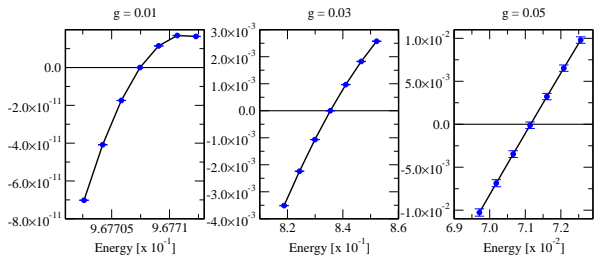}
\end{center}
\caption{\label{fig5} (Color online) 
Energy dependence of (the real part of) the 
function $Q(E, g)$. The results for the zeroes of 
$Q(E, g)$ obtained by a quadratic regression 
analysis are depicted by solid lines. Calculations have been performed 
around the "true" energies of the first excited state with 
$K$ = 1, $\varepsilon$ = +1 ($M$ = 1)
and for three different 
values of the coupling parameter: $g = 0.01$ (left panel), 
$g = 0.03$ (middle panel) and $g = 0.05$ (right panel).
The energy eigenvalues obtained using the displayed graphs determine the
corresponding entries in the right column of Table~\ref{table4}.}
\end{figure}

Until now we have discussed the computation of the ground--state 
energy $E^{(0)}_{\rm FP}$ of the Fokker--Planck Hamiltonian. Of course, the 
function $Q(E, g)$ may also help to determine the energies of 
excited states. In contrast to the ground state, however, the
numerical analysis of the function $Q(E, g)$ for excited states is more 
complicated due to bad convergence of the Borel sums
for the $A_{\rm FP}(E, g)$ and $B_{\rm FP}(E, g)$ functions
in the energy range relevant for the excited states.
As seen from Fig.~\ref{fig5}, the convergence problems lead 
to relatively large numerical 
uncertainties for the numerical calculation of the function 
$Q(E, g)$ already for a relatively mild coupling parameter 
$g = 0.07$. We recall that this value of $g$ corresponds to the 
minimum of the energy $E_{\rm FP}^{(M = 1)}(g)$ as a 
function of $g$ and thus can be naturally identified
as marking the transition from weak to strong coupling. 
As a result of the numerical
uncertainties, the energy $E_{\rm FP}^{(M = 1)}(g = 0.07)$ of the 
first excited state may be reproduced only up to 2 decimal digits 
(see Table~\ref{table4}). 
For even larger values of parameter $g$, the maximal accuracy of 
calculations, based on Eq.~(\ref{Q_zeros}),
is only a single significant digit, even though the 
double resummation of the instanton expansion, and of the
perturbative expansion about each instanton, is implicitly
contained in the cited Equation.

\begin{figure}[t]
\begin{center}
\includegraphics[height=0.9\linewidth,angle=270]{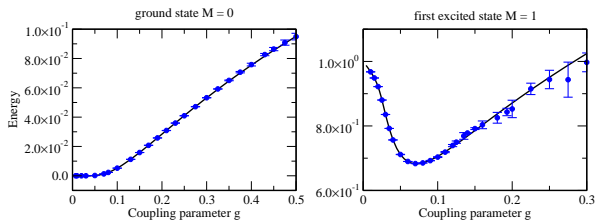}
\end{center}
\caption{\label{fig6} (Color online) Eigenvalues $E^{(M = 0)}_{\rm FP}$ (left panel) and 
$E^{(M = 1)}_{\rm FP} \equiv E^{(K = 1, \varepsilon = +1)}_{\rm FP}$ (right panel) 
of the Fokker--Planck Hamiltonian 
as a functions of the coupling parameter $g$. Results have been computed by the
diagonalization of the Fokker--Planck Hamiltonian in the basis of the harmonic 
oscillator wavefunctions (solid lines) and by solving Eq.~(\ref{Q_zeros}),
where the latter correspond the the data points with the error bars.}
\end{figure}

\begin{figure}[t]
\begin{center}
\includegraphics[width=0.6\linewidth,angle=270]{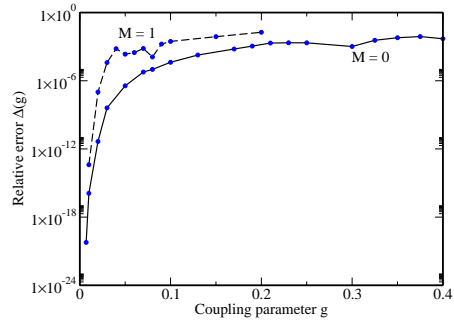}
\end{center}
\caption{\label{fig7} (Color online) Relative error (\ref{relative_error}) of 
calculations of the eigenvalues as a function of coupling constant $g$. 
Results are presented for the ground ($M$ = 0) and the first excited 
($M$ = 1) states of the Fokker--Planck potential.}
\end{figure}

Supplementing the the specific Fokker--Planck energies of the $M = 0,1$ states
presented in Tables~\ref{table3} and \ref{table4}, 
we indicate in Fig.~\ref{fig6} 
the $g$--dependence of the energies $E^{(0)}_{\rm FP}$ and $E^{(1)}_{\rm FP}$
as obtained from the analysis of the function $Q(E, g)$ and 
compare to reference values obtained from the 
diagonalization of the Hamiltonian matrix. As indicated in Fig.~\ref{fig6},
the results obtained by a direct resummation of the 
quantization condition give a good quantitative picture 
of the ground--state and the first excited--state eigenvalues in the 
ranges $0 \le g \le 0.7$ and $0 \le g \le 0.3$, respectively. 
The behavior of the numerical uncertainty as a function of $g$
is explicitly shown in Fig.~\ref{fig7}, where we plot the quantity
\begin{equation}
\label{relative_error}
\Delta(g) \, = \, 
\left| \frac{E_{\rm FP, \, resum}^{(M)}(g) - 
E_{\rm FP, \, diag}^{(M)}(g)}{E_{\rm FP, \, diag}^{(M)}(g)} 
\right| \, ,
\end{equation}
with $E^{(M)}_{\rm FP, \, resum}(g)$ and $E_{\rm FP, \, diag}^{(M)}(g)$ being 
the eigenvalues as obtained from the direct 
resummation of the quantization condition given in 
Eq.~(\ref{Q_zeros}) and the 
diagonalization of the Hamiltonian (\ref{hamfp}), 
respectively (the latter values, which are numerically 
more accurate, are taken as the reference values). While the accuracy of 
resummations for the ground--state energy remains 
satisfactory even for (relatively) strong coupling, the relative error 
for the first excited is numerically much more significant.

\begin{table}[t]
\caption{\label{table4} The energy of the first excited state 
with $M=1$
of the Fokker--Planck Hamiltonian. Results have been 
computed by the diagonalization of the Hamiltonian in the basis of the 
harmonic oscillator wavefunctions (exact energy) and by solving 
Eq.~(\ref{Q_zeros}).}
\vspace*{0.3cm}
\begin{scriptsize}
\begin{center}
\begin{tabular}{l@{\hspace*{0.5cm}}l@{\hspace*{0.5cm}}l} 
\hline
\hline
\rule[-3mm]{0mm}{8mm}
$g$   & $E^{(M = 1)}_{\rm FP}$ (diagonalization) & zero of the Re[$Q(E, g)$]  \\
\hline
\rule[-3mm]{0mm}{8mm}
0.010 &             9.677 074 461 352   $\times$ 10$^{-01}$  &
         \underline{9.677 074 461 352}  $\times$ 10$^{-01}$ \\
\rule[-3mm]{0mm}{8mm}
0.020 &             9.219 489 780 495   $\times$ 10$^{-01}$  &
         \underline{9.219 4}90(3)  $\times$ 10$^{-01}$ \\
\rule[-3mm]{0mm}{8mm}
0.030 &             8.354 795 860 905   $\times$ 10$^{-01}$  &
         \underline{8.354} 4(6)         $\times$ 10$^{-01}$ \\
\rule[-3mm]{0mm}{8mm}
0.070 &             6.828 548 309 058   $\times$ 10$^{-01}$  &
         \underline{6.8}33(4)           $\times$ 10$^{-01}$ \\
\rule[-3mm]{0mm}{8mm}
0.200 &             8.710 869 037 634   $\times$ 10$^{-01}$ & 
         \underline{8.7}6(5)            $\times$ 10$^{-01}$ \\
\rule[-3mm]{0mm}{8mm}
0.250 &             9.508 936 793 119   $\times$ 10$^{-01}$ & 
         \underline{9}.4(2)             $\times$ 10$^{-01}$ \\
\hline
\hline
\end{tabular}
\end{center}
\end{scriptsize}
\end{table}

%
%
\subsection{Strong--coupling expansion}
\label{largec}

As discussed in the previous Section, a numerical procedure based on
the generalized Bohr--Sommerfeld formulae may provide relatively accurate
estimates of the ground as well as the (first
two) excited--state energies for the 
coupling parameters in the range $0 \le g \le 0.3$. 
The question arises whether $g \approx 0.3$ can be considered as 
belonging to the strong coupling regime. Since the minima of the 
first two excited-state energies occur at $g = 0.07$ and $g = 0.025$,
respectively, one might be tempted to answer the question affirmatively.
However, one could devise a different
criterion for the transition to the strong
coupling regime. For instance, one might define the strong--coupling region 
as a region of an appropriately specified
large-coupling asymptotic behavior of the eigenvalues 
$E_{\rm FP}^{(M)}(g)$. 

The large--coupling asymptotics of the Fokker--Planck potential thus 
represents a natural next aim in the current investigation. 
To this end, we apply a
so--called Symanzik scaling $q \rightarrow g^{-1/6} q$ in Eq.~(\ref{hamfp}) 
and rewrite the Fokker--Planck potential into another one 
with the same eigenvalues but a fundamentally different structure 
\cite{SiDi1970, We1996b}: 
\begin{eqnarray}
\label{scaled_hamiltonian}
H_{\rm FP} = g^{1/3} 
\left[ H_{\rm L} + 
\left( -q^3 - \frac{1}{2}\right)g^{-1/3} + \frac{q^2}{2} g^{-2/3}
\right] \, ,
\end{eqnarray}
where the Hamiltonian $H_{\rm L}$ does not depend on $g$:
\begin{eqnarray}
\label{HL}
H_{\rm L} = -\frac{1}{2} \left( \frac{d}{dq} \right)^2 + 
q + \frac{q^4}{2} \, .
\end{eqnarray}
We conclude that the $M$th eigenvalue of the Fokker--Planck Hamiltonian
for the $g \rightarrow \infty$ is determined in leading order by the 
$M$th eigenvalue $E_{\rm L}^{(M)}$ of the Hamiltonian (\ref{HL}):
\begin{equation}
   \label{leading_asymptotics}
   E_{\rm FP}^{(M)}(g) \approx g^{1/3} E_{\rm L}^{(M)} \, .
\end{equation}
Moreover, Eq.~(\ref{leading_asymptotics}) also indicates that 
the classifications of the levels $M = 0, 1, 2, \dots$ of the
Fokker--Planck and the $H_{\rm L}$ Hamiltonians are obviously identical 
in the strong--coupling regime.

\begin{figure}[t]
\begin{center}
\includegraphics[width=0.8\linewidth,angle=270]{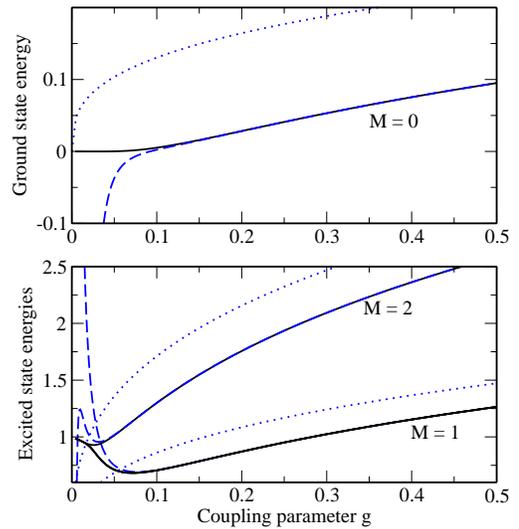}
\end{center}
\caption{\label{fig8} (Color online)
Exact values (solid line) for the ground state 
and the two lowest excited energy levels 
for the Fokker--Planck potential as a function of $g$, together with the
leading asymptotics (dotted line) and the  
partial sum of the strong-coupling expansion (dashed line) as defined by the 
first six nonvanishing terms in powers of $ g^{-2k/3}$
[see Eq.~(\ref{strong_coupling_expansion})], which are
listed in Table~\ref{table3}. Note that the leading strong-coupling 
asymptotics alone cannot satisfactorily 
describe the true energy eigenvalues for 
moderate and small coupling. By contrast, the partial sum
of the first six nonvanishing terms 
of the strong-coupling asymptotics yields numerical values
which are indistinguishable from the true eigenvalues on the level 
of the line thickness used in the plots, down to rather small
values of the coupling (dashed vs.~solid lines).}
\end{figure}

Based on Eq.~(\ref{leading_asymptotics}), we now wish to compute the leading
asymptotics of the ground $E_{\rm FP}^{(M = 0)}(g)$ and excited--state 
$E_{\rm FP}^{(M = 1, 2)}(g)$ energies of the Fokker--Planck
Hamiltonian. This computation obviously requires information about the
corresponding eigenvalues of the Hamiltonian $H_{\rm L}$. The
energies $E_{\rm L}^{(M)}$ have again been determined by a
diagonalization of the 
Hamiltonian matrix within a basis of up to 1000 harmonic oscillator
wavefunctions and than utilized in Eq.~(\ref{leading_asymptotics}). 
As seen from Fig.~\ref{fig8}, the leading asymptotics of the eigenvalues 
$E_{\rm FP}^{(M = 0, 1, 2)}(g)$ (dotted line), 
calculated in such a way, significantly 
overestimate the energies of the Fokker--Planck Hamiltonian for the
region $0 \le g \le 0.3$. Higher--order corrections
to the large-coupling asymptotics are therefore required,
in order to reproduce more accurately the asymptotics of the eigenvalues
$E_{\rm FP}^{(M = 0, 1, 2)}(g)$. We observe that 
we may apply standard Rayleigh--Schr\"{o}dinger perturbation theory 
to Eq.~(\ref{scaled_hamiltonian}) and use the fact that 
the perturbative with respect to $H_{\rm L}$, which is 
$V(g) = \left( -q^3 - 1/2 \right)g^{-1/3} + q^2 g^{-2/3}/2$,
remains Kato--bounded
with respect to the unperturbed Hamiltonian $H_{\rm L}$ for large $g$.
Within such an approach, a strong--coupling perturbation expansion can be
written for each energy $E_{\rm FP}^{(M)}(g)$:
\begin{equation}
   \label{strong_coupling_expansion}
   E_{\rm FP}^{(M)}(g) = g^{1/3} \sum\limits_{k = 0}^{\infty} L^{(M)}_k 
   g^{-2k/3} \,
\end{equation}
where $L^{(M)}_0 \equiv E_{\rm L}^{(M)}$ and the higher perturbation
coefficients $L^{(M)}_{k > 0}$ are calculated in the basis of the wavefunctions
of the unperturbed Hamiltonian (\ref{HL}). The first six $L^{(M)}_{k}$
coefficients for the ground $M$ = 0 as well as the first excited  $M$ = 1, 2
states are given in Table~\ref{table5}. 

In Fig.~\ref{fig8}, we implemented the first few 
expansion coefficients as listed in Table~\ref{table5}, in order to calculate 
strong-coupling asymptotics 
for the three lowest levels of the Fokker--Planck Hamiltonian 
(dashed lines). Obviously, the 
partial sum of the expansion (\ref{strong_coupling_expansion}) as defined by
only six coefficients provides much better agreement with 
numerically determined energy eigenvalues 
than the leading asymptotics (\ref{leading_asymptotics}). For instance, 
the energy of the first excited state $M = 1$ is very well described by the
(first six terms of the) strong--coupling expansion already at $g = 0.07$
(the agreement is better than one percent). 
This point, which
also corresponds to the minimum of the function $E_{\rm FP}^{(M = 1)}(g)$, 
can thus naturally be identified as the transition region
between the regimes of weak and strong coupling. By using such a 
definition for the transitory regime, we
may finally conclude that the resummation of the Bohr--Sommerfeld quantization
formulae (\ref{qcond})--(\ref{A_expansion}),
using the condition (\ref{Q_zeros}), may provide reasonable estimates
for the energy levels $E_{\rm FP}^{(M)}(g)$ even in a limited 
subregion of the strong-coupling regime.

%
%
\subsection{Quantum dynamics}
\label{quadyn}

\begin{figure}[t]
\begin{center}
\includegraphics[width=0.5\linewidth,angle=270]{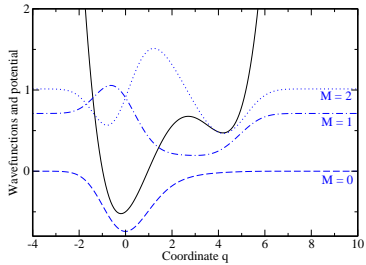}
\end{center}
\caption{\label{fig9} (Color online)
The wavefunctions $\Phi^{(M)}_{\rm FP}(q)$ of the ground $M$ = 0 and the first
excited $M$ = 1, 2 states of the Fokker--Planck Hamiltonian calculated at
coupling parameter $g$ = 0.05. The base lines for the plots of the wave 
functions correspond to their energies $E_{\rm FP}^{(M)}(g)$.}
\end{figure}

In previous Sections of the current paper, 
we have presented a systematic study of the energy
spectrum of the Fokker--Planck potential. In particular, two methods for
computing the eigenvalues $E_{\rm FP}^{(M)}(g)$ 
have been discussed in detail: (i) 
a ``brute--force'' method which is based on the diagonalization of the 
Hamiltonian matrix and (ii) a generalized perturbative approach which 
accounts for the instanton effects.  
While, of course, the precise computation of the energy levels is 
a very important
task, the complete description of the properties of the particular 
Hamiltonian also requires an access to its wavefunctions 
$\Phi^{(M)}_{\rm FP}(q)$. Most naturally, these wavefunctions may be obtained
together with the eigenvalues $E_{\rm FP}^{(M)}(g)$ by 
matrix diagonalization.
In our calculations, the basis of the standard harmonic
oscillator wavefunction $\{ \phi_n(q) \}_{n=0}^{\infty}$ is used for
the construction of a Hamiltonian matrix 
$\mem{\phi_n}{H_{\rm FP}}{\phi_m}$. The 
Fokker--Planck eigenfunctions are thus given by:
\begin{equation}
   \label{FP_wavefunctions}
   \Phi^{(M)}_{\rm FP}(q) = 
   \sum\limits_{n = 0}^{\infty} c^{(M)}_n \phi_n(q) \, ,
\end{equation}
where the coefficients $c^{(M)}_n$ are found by the diagonalization procedure.
In Fig.~\ref{fig9} we display, for example, the wavefunctions 
(\ref{FP_wavefunctions}) of the ground $M$ = 0 and the first excited 
$M = 1, 2$ states as calculated for a coupling parameter $g = 0.05$. 
As is evident from Fig.~\ref{fig9}, the 
symmetry--breaking term of the Hamiltonian 
(\ref{hamfp}) leads to a ground--state wavefunction 
$\Phi^{(M = 0)}_{\rm FP}(q)$ (dashed line) which is neither a symmetric nor 
antisymmetric combination of the wavefunctions of the right and left wells,
but localized in the lower well. For the first
excited states, in contrast, the symmetry is partially restored, and we may
attribute the wavefunctions $\Phi^{(M = 1)}_{\rm FP}(q)$ (dashed and dotted 
line) and $\Phi^{(M = 2)}_{\rm FP}(q)$ (dotted line) to states with 
positive ($\varepsilon$ = +1) and negative ($\varepsilon$ = -1) parities 
[see also Eqs.~(\ref{FP_M_to_K})---(\ref{FP_M_to_epsilon})].

%
%
\begin{center}
\begin{table}
\begin{center}
\caption{\label{table5} $L^{(M)}_{k}$ coefficients of the strong--coupling
perturbation expansion (\ref{strong_coupling_expansion}) of the eigenvalues
$E_{\rm FP}^{(M = 0, 1, 2)}$ of the Fokker--Planck Hamiltonian.}
\begin{tabular}{c@{\hspace{0.6cm}}r@{\hspace{0.3cm}}r@{\hspace{0.3cm}}r}
\hline
\hline
$j$ &
\multicolumn{1}{c}{$M=0$} &
\multicolumn{1}{c}{$M=1$} &
\multicolumn{1}{c}{$M=2$} \\
\hline
0 &  $0.281\,067\,805$ &  $1.854\,587\,292$ &  $3.686\,419\,624$ \\
1 & $-0.132\,985\,313$ & $-0.209\,853\,650$ & $-0.307\,985\,031$ \\
2 &  $0.021\,367\,333$ &  $0.028\,174\,214$ &  $0.025\,765\,237$ \\
3 & $-0.000\,876\,935$ &  $0.000\,220\,875$ & $-0.000\,099\,791$ \\
4 & $-0.000\,060\,335$ &  $0.000\,031\,789$ & $-0.000\,002\,135$ \\
5 & $-0.000\,001\,557$ &  $0.000\,000\,739$ & $-0.000\,000\,385$ \\
\hline
\hline
\end{tabular}
\end{center}
\end{table}
\end{center}
%
%

Until now, we have only discussed the evaluation of the eigenstates
and eigenenergies 
of the Fokker--Planck Hamiltonian. In theoretical studies of 
double quantum-dot nanostructures \cite{GrDiJuHa1991, Op1999, GrHa1998, TsOp2004}
and of the quantum tunneling phenomena in atomic physics
\cite{LiBa1990, LiBa1992}, a large number of problems arise, 
in which the time-propagation of some (specially prepared) wavepacket in 
double-well-like potentials has to be considered. The methods
for such a time-propagation, such as the well-known Crank-Nicolson method, 
the split-operator technique, and approaches based on the Floquet 
formalism and many others, are discussed in detail 
in the literature (see, e.g., 
Refs.~\onlinecite{GrDiJuHa1991, GrHa1998,CrNi1947, FlMoFe1976, PrKeKn1997}). 
As a supplement to our previous considerations,
we will now consider the time-propagation of an initial wavepacket in a
double-well-like potential, recalling the adiabatic
approach as one of the most simple and best-known techniques
(see, e.g., the book~\onlinecite{Te2003} and references therein) 
for the integration
of the (time--dependent) single particle Schr\"odinger equation.
Within such a technique, in which we can naturally make use of the 
results previously derived for the eigenstates
and eigenenergies, the propagation of a wavepacket $\Psi(q, t)$ in 
the (time--independent) Fokker--Planck potential (\ref{hamfp}) is 
given by:
\begin{equation}
\label{propagation_FP}
\Psi(q, t) = \sum\limits_{M=0}^{\infty} 
b^{(M)} \,
\exp\left(-{\rm i}\, E^{(M)}_{\rm FP} t\right) \, 
\Phi^{(M)}_{\rm FP}(q) \,.
\end{equation}
Here the coefficients $b^{(M)} = \sprm{\Phi^{(M)}_{\rm FP}}{\Psi(t=0)}$
determine the decomposition of the initial wavepacket (at $t$ = 0) in the basis
of the eigenfunctions (\ref{FP_wavefunctions}).

Equation~(\ref{propagation_FP}) provides an exact solution for
the wavefunction $\Psi(q, t)$ at an arbitrary moment of time only 
in the limit of an infinitely large basis of harmonic oscillator 
$\{ \phi_n(q) \}_{n=0}^{\infty}$ and Fokker--Planck 
$\{ \Phi^{(M)}_{\rm FP}(q) \}_{M=0}^{\infty}$ wavefunctions. For computational
reasons, however, summations over basis functions have to be 
restricted to a finite number. In our calculations,
basis sets of 300--1000 wavefunctions have been applied depending on the
parameters of the initial wavepacket and the coupling parameter $g$. The
actual size of the basis has been chosen according to the numerical checks of
the Ehrenfest theorem or the normalization of the wavepacket.    

\begin{figure}[t]
\begin{center}
\includegraphics[width=0.7\linewidth,angle=270]{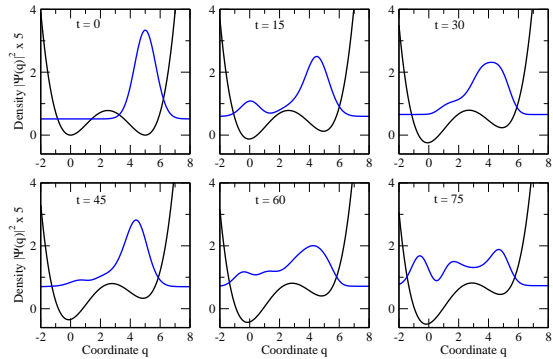}
\end{center}
\caption{\label{fig10} (Color online)
Time propagation of the (initially) Gaussian wavepacket in the time--dependent
potential (\ref{hami}) with $\eta = \sin(t/60)$ which oscillates between the 
double--well and Fokker--Planck potentials. The base line
for the plot of the wave function corresponds to its instantaneous 
energy.}
\end{figure}

Within the adiabatic approximation, which is valid for 
slowly varying potentials~\cite{Te2003,CN}, we may divide the 
time evolution of the potential into small intervals $\Delta t$ and assume that
for every $k$th interval the Hamiltonian is time--independent and the 
propagation of the wavepacket is governed again by Eq.~(\ref{propagation_FP})
where, of course, eigenvalues $E^{(M)}_{\rm FP}$ and 
eigenfunctions $\Phi^{(M)}_{\rm FP}(q)$ should be replaced with the eigenvalues
$E^{(M)}_{k}$ and eigenfunctions $\Phi^{(M)}_{k}(q)$ of the 
Hamiltonian $H_k \equiv H(t_k)$. We have applied this adiabatic 
time--propagation method, 
whose variations are well known from the 
literature \cite{HwPe1977,Ve2000,Su2003,Te2003} and which is 
equivalent to an exponentiation of the instantaneous
Hamiltonian for each time interval $\Delta t$,
to investigate the evolution of an (initially) Gaussian wavepacket in 
a time--dependent potential (\ref{hami}) which oscillates 
sinusoidally between the Fokker--Planck and the 
double-well cases. Since the animated results of this simulation are 
available for download\cite{JeHomeHD}, we just present a
small series of snapshots in Fig.~\ref{fig10}. 
As seen from these pictures, the wavepacket, which 
is initially located in the right well, performs 
oscillations between the wells.
These oscillations are controlled by the temporal change of 
the potential (\ref{hami}). We have checked empirically that the adiabatic
approximation employed here does not represent an obstacle for an accurate
time evolution in even rapidly oscillating potentials,
because of the calculational efficiency of the other steps in our time
propagation algorithm (notably, the diagonalization including the 
determination of eigenvectors can be implemented 
in a computationally very favorable way on modern computers).
It is thus possible to perform quantum dynamical simulations 
in potentials which oscillate between two limiting forms with 
two fundamentally different characteristic ground-state configurations,
each of which is governed by instantons, though in a different way.
A generalization of our approach to two-dimensional potentials
appears to be feasible and is currently being studied.

%
%
\section{Conclusions}
\label{conclu}

In this paper, 
we have investigated the Fokker--Planck [Eq.~(\ref{hamfp})]
and the double-well [Eq.~(\ref{hamdw})] potential from the point of 
view of large-order perturbation theory (resurgent expansion and 
generalized quantization condition), in order to map out the regimes 
of validity of the instanton-related resurgent expansion for the 
lowest energy levels, and in order to explore the possibility of 
reaching the strong-coupling regime via a direct resummation
of the $A_{\rm FP}$ and $B_{\rm FP}$ functions
given in Eqs.~(\ref{B_expansion}) and (\ref{A_expansion}), which enter the 
generalized quantization condition (\ref{qcond}).
The latter approach entails a complete double resummation
of the resurgent expansions (\ref{eground}) and 
(\ref{efpgen}) both in powers of the 
instanton coupling $\exp(-1/6g)$ and in powers of the coupling $g$
(perturbative expansion about each instanton). The quest has been
to explore the applicability of resummed expansions for medium
and large coupling parameters, in the transitory regime to large coupling.

It is quite natural to identify the transition region for the 
first excited state as defined 
by the minimum of the energy level $E_{\rm FP}^{(M=1)}(g)$ as a 
function of $g$ (see the right panel of Fig.~\ref{fig6}),
which occurs near $g \approx 0.07$.
As is evident from Figs.~\ref{fig6} and~\ref{fig8}, it is possible to reach 
convergence for both the resummed secular equation (\ref{Q_zeros}) 
as well as convergence of the strong-coupling 
expansion~(\ref{strong_coupling_expansion}) 
in a somewhat restricted overlap region $0.04 \lesssim g \lesssim 0.3$.
The question whether it is possible to use the resummed instanton
expansion for large coupling, cannot universally be answered affirmatively, 
although it is reassuring to find at 
least a restricted region of overlap. In order to interpret
the overlap, one should remember that for 
typical Borel summable series
as they originate in various contexts in field theory,
it is much easier to perform resummations at moderate and even large
coupling parameters than for the instanton-related case considered
here. An example is the 30-loop resummation
of the anomalous dimension $\gamma$ function of six-dimensional
$\phi^3$-theories and of Yukawa model theories~\cite{BrKr2000},
which lead to excellent convergence for couplings as large
as $g = 10$ and higher~\cite{JeSo2001}. 

For the latter case, it is even possible to obtain, numerically,
the strong-coupling asymptotics on the basis of a resummed
weak-coupling perturbation theory (see, e.g., Ref.~\onlinecite{Su2001}
for a remarkable realization of this idea in an 
extremely nontrivial context). The general notion is that any perturbation
(a potential in the case of quantum mechanics and 
an interaction Lagrangian in the case of field theory)
determines the large-order behavior of perturbation series
describing a specific physical quantity.
However, the potential or interaction Lagrangian 
also determine the large-coupling expansion for the 
physical quantity under investigation.
This means that there is a connection between the large-coupling expansion
and the large-order behavior of the perturbation series generated
in each theory, and this correspondence can be exploited in order
to infer strong-coupling asymptotics even in cases where only 
a few perturbative terms are known~\cite{Su2001}.
According to our numerical 
results, the corresponding calculation of strong-coupling asymptotics
on the basis of only a few perturbative terms would be much more difficult in
cases where instantons are present, even if like in our case,
additional information is present in the form of the generalized quantization
condition (\ref{qcond}).

We deliberately refrain from speculating about further implications of this
observation and continue with a summary of the application-oriented 
results gained in the current investigation.  
Potentials of the double-well type are important for a number
of application--oriented calculations, including
Josephson junction qubits~\cite{BeEtAl2003, WeEtAl2005}, 
inversion doubling in molecular physics \cite{Hu1927,Xu2005},
semiconductor double 
quantum dots~\cite{LoDV1998, HoEtAl2002, HuEtAl2005},
as well as, in a wider context, Bose--Einstein condensates in 
multi--well traps~\cite{StCeMo2004, AlStCe2005, StAlCe2006, AlEtAl2005}.
For the latter case, 
a number of theoretical works have been performed recently in 
order to explore the time propagation of (one particle) 
wavepackets in driven double--well potentials
\cite{GrDiJuHa1991,Op1999,TsOp2004}. In Sec.~\ref{numer}
we discuss a numerical procedure for an accurate description of 
energy levels and of the corresponding wave functions, which can thus be 
used in order to construct basis sets for an accurate quantum dynamical time
evolution of wave packets in both static
as well as time--dependent potentials. This well-known adiabatic technique 
for the integration of the (single particle) time-dependent 
Schr\"odinger equation is briefly recalled in Sec.~\ref{quadyn}.
We illustrate this technique by a calculation for a potential
which oscillates between the Fokker--Planck and the double-well
cases, governed by a time-dependent
interpolating parameter $\eta = \cos(\omega t)$ 
as given in Eq.~(\ref{hami}). Similar calculations can be done for 
cases where the potential admits resonances as in the 
case of a cubic anharmonic oscillator. In this case, the 
method of complex scaling leads to a basis of states which can be used 
in order to start quantum dynamical simulations. 
Related work is currently in progress.

%
%
\section{Acknowledgments}

U.D.J.~acknowledges support from the Deutsche
Forschungsgemeinschaft (Heisenberg program).
M.L.~is grateful to Max--Planck--Institute for Nuclear Physics
for the stimulating atmosphere during a guest researcher
appointment, on the occasion of which the current work was
completed. 

\appendix

%
%
\section{Supersymmetry and the Fokker--Planck Potential}
\label{appa}

This appendix is meant to provide a brief identification 
of the Fokker--Planck potential (\ref{hamfp}) in terms
of a supersymmetric (SUSY) algebra~\cite{CoKhSu1995,KaSo1997}. 
We define the operators
\begin{equation}
B^\pm = \frac{1}{\sqrt{2}} \, 
\left( W(q) \mp \frac{d}{dq} \right)
\end{equation}
with $W(q) = q \, (1- \sqrt{g} \, q)$ and the SUSY Hamiltonian 
\begin{equation}
H_{\rm SUSY} = 
\left( \begin{array}{cc} B^+ B^- & 0 \\ 0 & B^- B^+ \end{array} \right) =
\left( \begin{array}{cc} H_1 & 0 \\ 0 & H_2 \end{array} \right) \,.
\end{equation}
with 
\begin{subequations}
\begin{align}
H_1 =& -\frac{1}{2} \left( \frac{d}{dq} \right)^2 + 
\frac{1}{2} q^2 \left( 1 - \sqrt{g}q \right)^2 + 
\sqrt{g} q - \frac{1}{2} \,, \\
H_2 =& -\frac{1}{2} \left( \frac{d}{dq} \right)^2 + 
\frac{1}{2} q^2 \left( 1 - \sqrt{g}q \right)^2 - 
\sqrt{g} q + \frac{1}{2} \,.
\end{align}
\end{subequations}
Notice that $W(q)$ finds a natural interpretation as a 
``superpotential'' in the sense of Refs.~\onlinecite{CoKhSu1995,KaSo1997}.
The Hamiltonians $H_1$ and $H_2$ are ``superpartners.''
They are related to each other by a simple reflection and 
translation, $q \to 1/\sqrt{g} - q$, and have the same 
spectra. In that sense, one may say that the Fokker--Planck 
potential is its own superpartner. The Fokker--Planck potential
therefore constitutes 
a case of ``broken supersymmetry'' with zero Witten index
[see, e.g., Eq.~(2.88) of Ref.~\onlinecite{KaSo1997}].
The construction of the supersymmetric partner thus does not 
help in the analysis of the Fokker--Planck Hamiltonian. One is 
forced into the instanton-inspired analysis presented in the current 
study. 

As a last remark, we recall that the Fokker--Planck potential
\begin{equation}
V_{\rm FP}(q) = \frac{1}{2} q^2 \left( 1 - \sqrt{g}q \right)^2 + 
\sqrt{g} q - \frac{1}{2} = \frac12\, \left[W^2(q) - W'(q) \right] 
\end{equation}
could be assumed to admit a zero eigenvalue. However, 
as is evident from the 
discussion following Eq.~(7.28) of Ref.~\onlinecite{ZJJe2004ii},
the corresponding eigenfunction is not normalizable, 
and thus cannot be interpreted as a physical state vector.
Neither the Fokker--Planck potential nor its isospectral
supersymmetric partner admit a zero eigenvalue.

\end{document}